\documentclass[12pt]{article}
\usepackage{amsmath}
\usepackage{times}
\usepackage{graphicx}
\usepackage{multirow}
\usepackage{theorem}
\usepackage[authoryear]{natbib}
\usepackage{rotating}
\usepackage{bbm}
\usepackage{latexsym}
\usepackage{setspace}

\theorembodyfont{\itshape}
\newtheorem{Theorem}{Theorem}
\theorembodyfont{\itshape}
\newtheorem{Lemma}[Theorem]{Lemma}
\theorembodyfont{\rmfamily}
\newtheorem{Definition}[Theorem]{Definition}
\theorembodyfont{\rmfamily}
\newtheorem{Proof}{Proof:}

\newtheorem{Example}[Theorem]{Example}

\newcommand{\captionfonts}{\normalsize}

\makeatletter  
\long\def\@makecaption#1#2{%
  \vskip\abovecaptionskip
  \sbox\@tempboxa{{\captionfonts #1: #2}}%
  \ifdim \wd\@tempboxa >\hsize
    {\captionfonts #1: #2\par}
  \else
    \hbox to\hsize{\hfil\box\@tempboxa\hfil}%
  \fi
  \vskip\belowcaptionskip}
\makeatother   

\begin{document}
\hspace{13.9cm}1

\ \vspace{20mm}\\

{\LARGE On the Relation between Encoding and Decoding of Neuronal Spikes}

\ \\
{\bf \large Shinsuke Koyama}\\
{Department of Mathematical Analysis and Statistical Inference, 
Institute of Statistical Mathematics, 
Tokyo, 190-8562, Japan}\\
%


\thispagestyle{empty}
\markboth{}{NC instructions}
\ \vspace{-0mm}\\
%
\begin{center} {\bf Abstract} \end{center}
Neural coding is a field of study that concerns how sensory information is represented in the brain by networks of neurons. 
The link between external stimulus and neural response can be studied from two parallel points of view. 
The first, neural encoding refers to the mapping from stimulus to response, 
and primarily focuses on understanding how neurons respond to a wide variety of stimuli, and on constructing models that accurately describe the stimulus-response relationship. 
Neural decoding, on the other hand, refers to the reverse mapping, from response to stimulus, 
where the challenge is to reconstruct a stimulus from the spikes it evokes.
Since neuronal response is stochastic, a one-to-one mapping of stimuli into neural responses does not exist, causing a mismatch between the two viewpoints of neural coding. 
Here, we use these two perspectives to investigate the question of what rate coding is,  
in the simple setting of a single stationary stimulus parameter and a single stationary spike train  represented by a renewal process. 
We show that when rate codes are defined in terms of encoding, i.e., the stimulus parameter is mapped onto the mean firing rate, the rate decoder given by 
spike counts or the sample mean, does not always efficiently decode the rate codes, but can improve efficiency in reading certain rate codes, when correlations within a spike train are taken into account.

\section{Introduction}
Sensory and behavioral states are represented by neuronal responses. Determining which code is used by neurons is important in order to understand how the brain carries out information processing \citep{Dayan-Abbott01,Rieke-et-al97}.
Coding schemes used by neurons can be divided approximately into two categories. 
In rate coding,  
the stimulus is mapped onto the firing rate, defined as the average number of spikes per unit time. 
A variation in the number of emitted spikes in response to the same stimulus across trials, is then considered noise. 
In temporal coding, on the other hand, 
the stimulus is encoded in moments of the spike pattern that have higher order than the mean \citep{Theunissen95}.

While neural codes are characterized in terms of these encoding views 
(i.e., how the neurons map the stimulus onto the features of spike responses), 
these are often investigated and validated using decoding. 
From the decoding viewpoint, rate coding is operationally defined by counting the number of spikes over a period of time, without taking into account any correlation structure among spikes.
Any scheme based on such an operation is equivalent to decoding under the stationary Poisson assumption, because the number of spikes over a period of time, or the sample mean of interspike intervals (ISIs), is a sufficient statistic for the rate parameter of a homogeneous Poisson process. 
In this manuscript, a decoder based on counting the number of spikes, or on taking the sample mean of ISIs, is labeled as ``rate decoder".
Similarly, temporal coding can be defined by decoding the stimulus using a statistical model with a correlation structure between spikes (such as the MI model, introduced below). If such a decoder improves on the performance of the rate decoder, it indicates that significant information about the stimulus is carried in the temporal aspect of spike trains \citep{Jacobs-et-al09, Pillow05}.

A simple statistical model with a correlation structure has been introduced in the literature, 
taking the intensity function of a point process to be a product of two factors:
\begin{equation}
\label{eq:mi}
\lambda(t, s_*(t)) = \phi(t)g(t-s_*(t)),
\end{equation}
where $s_*(t)$ represents the last spike time preceding $t$.
This statistical model with the intensity function (\ref{eq:mi}) has been called the multiplicative intensity (MI) model by \cite{Aalen78} and the multiplicative inhomogeneous Markov interval model by \cite{Kass-Ventura01}. 
$\phi(t)$ is the free firing rate, which depends only on the stimulus, 
and $g(t-s_*(t))$ is the recovery function, which describes the dependency of the last spike time preceding $t$ and hence allows the MI model to have a correlation structure between spikes.  
Note that Eq.(\ref{eq:mi}) becomes the intensity function of an inhomogeneous Poisson process if the recovery function is constant in time. 
It has been reported that the MI model enhances decoding performance in real data analysis \citep{Jacobs-et-al09}, which encourages use of the MI model to test temporal codes.

Although neural codes can be defined in terms of either encoding or decoding, the resulting codes generally differ from one another. 
Here, we investigate the relation between the two viewpoints of neural coding in terms of rate and temporal coding schemes. 
Specifically, we consider, for the sake of analytical tractability, a simple setting of a single stationary stimulus parameter and a single stationary spike train represented by a renewal process, and investigate the extent to which decoders of each scheme decode neural codes that are defined in terms of encoding. 
Our main claim is that when rate codes are defined in terms of encoding, i.e., the stimulus parameter is mapped onto the mean firing rate, the rate decoder does not always efficiently decode the rate codes, whereas the temporal decoder can improve efficiency in reading certain rate codes.

In order to deduce our results, we develop, in section \ref{sec:theory}, a statistical theory based on asymptotic estimation, i.e., inference from a large number of ISIs. 
However, care must be taken when results based on asymptotic analysis are translated into non-asymptotic cases, which are certainly relevant in more realistic coding contexts.
This will be addressed in section \ref{sec:discussion}.

\section{Theory}\label{sec:theory}
\subsection{Definition of encoding and decoding}

We suppose, for simplicity, that neural spikes are described by a stationary renewal process. 
The response of single neurons is then described by an ISI density, 
$p(x|\theta)$, where $x \in [0, \infty)$, and $\theta \in \Theta \subset (-\infty,\infty)$ is a one-dimensional stimulus parameter. 
The renewal assumption is not exactly true for actual neural data, but often provides a reasonable approximation \citep{Troy92}.
Let $\mu=E(x|\theta)$ be the mean parameter, $E(\cdot|\theta)$ being the expectation with respect to $p(x|\theta)$. 

Consider first the rate encoding scheme. 
Since the early work of \cite{Adrian26}, 
there has been a search for a functional relationship 
between stimulus parameters and the average firing rate, which is often described as a function of the stimulus parameters.
This motivates us to formulate 
rate encoding as a one-to-one mapping from $\theta$ to $\mu(\theta)$. The variation in $x$ around the mean $\mu$ is then regarded as noise. 
In short, the rate encoding scheme can formally be defined as follows:

\begin{Definition}
If there exists a one-to-one and differentiable mapping $\theta \mapsto \mu(\theta)$, the scheme is rate encoding. 
\end{Definition}
The assumption of differentiability in $\mu(\theta)$ with respect to $\theta$ is required for analytical purposes, but is also reasonable physiologically because it shows that a small change in $\theta$ results in a small, smooth change in $\mu(\theta)$. 

Temporal encoding, on the other hand, intuitively means that the stimulus is encoded in statistical structures of ISIs beyond the firing rate. 
Since it allows for many alternatives, we do not explicitly define temporal encoding here, but instead give an example below.
Let $p(x|\mu,\kappa)$ be a dispersion model, where $\mu$ is the mean and $\kappa$ is the dispersion parameter that characterizes moments of the ISIs of higher order than the mean.
If the stimulus parameter is mapped onto the dispersion parameter, $\theta \mapsto \kappa(\theta)$, this scheme can be categorized under temporal encoding \citep{Kostal11}. 

For decoding, we assume an ISI density, $q(x|\phi)$, $\phi \in \Phi \subset (-\infty,\infty)$, which is chosen according to the decoding schemes introduced below. 
We suppose that decoding is performed by the maximum likelihood estimation (MLE) with $q(x|\phi)$. 
In rate decoding, 
one usually counts the number of spikes over a period of time, without taking into account any dependency among spikes. 
This is equivalent to decoding under the Poisson assumption, because the number of spikes is a sufficient statistic for the rate parameter of a homogeneous Poisson process. Thus, $q(x|\phi)$ is taken to be the exponential distribution, $q(x|\phi)=\phi\exp(-\phi x)$, for rate decoding. 

In temporal decoding, on the other hand, where a temporal dependency of spike timing relative to the last spike is considered, we take $q(x|\phi)$ to be the MI model.
Here, the ISI distribution of the MI model is constructed as follows. 
Since we only take into account stationary renewal processes, 
the rate factor in Eq.(\ref{eq:mi}) is reduced to a constant, and then
the intensity function, $\lambda(x)$, of the MI model becomes
\begin{equation*}
\lambda(x) = \phi g(x),
\end{equation*}
where $\phi \in [0,\infty)$ is the free firing rate and $g(x) (\ge 0)$ is the recovery function 
\footnote{
Since the units of $\lambda(x)$ are those of firing rate (i.e., spikes per unit time), by convention, we let $\phi$ also have units of firing rate, leaving $g(x)$ dimensionless \citep{Kass-Ventura01}.
}. 
The ISI distribution of the MI model is then obtained as
\begin{equation}
\label{eq:mimodel}
q(x|\phi) = \phi g(x) \exp[-\phi G(x)],
\end{equation}
where 
\begin{equation*}
G(x) = \int_0^x g(u)du.
\end{equation*}
In order for the MI model to be well behaved as a decoder, we assume that the variance of $G(x)$ is finite.
It is obvious from the factorization theorem \citep{Schervish95} that $G(x)$ is a sufficient statistic for $\phi$. 
Note that Eq.(\ref{eq:mimodel}) becomes an exponential distribution with firing rate $\phi$ if $g(x)=1$, $x\ge 0$. 
The two decoding schemes are summarized as follows:
\begin{Definition}
In rate decoding, $\theta$ is decoded with $q(x|\phi)$ being the exponential distribution via the MLE.
In temporal decoding, $\theta$ is decoded with $q(x|\phi)$ being the MI model via the MLE.
\end{Definition}

We use the MI model in temporal decoding for the following reasons. 
First, the inhomogeneous version of the MI model given by Eq.(\ref{eq:mi}) is useful in practice, as it can be easily fitted to data by well-established statistical methods \citep{Kass-Ventura01,DiMatteo01}. 
In fact, \cite{Jacobs-et-al09} demonstrated the importance of temporal coding by using this model. 
Second, generalized linear models (GLMs) \citep{McCullagh89,Paninski04,Paninski-et-al07,Truccolo05}, which have  been used extensively for statistical analysis of neural data, include the MI model as a special case. 
Specifically, the GLM corresponds to the MI model when the spiking history term contains only the last spike and a log-link function is used (e.g., soft-threshold integrate-and-fire models \citep{Paninski-et-al08}).

In order to investigate the extent to which decoders of each scheme decode neural codes that are defined in terms of encoding,
in section \ref{sec:correlation}, we introduce a correlation quantity $\rho^2_{\theta}$ given by Eq.(\ref{eq:rho2}), which measures decoding performance with $q(x|\phi)$.

\subsection{Correlation quantity}
\label{sec:correlation}

We shall assume that $p(x|\theta)$ and $q(x|\phi)$ satisfy the traditional regularity assumptions needed for standard asymptotics \citep{Schervish95}.
We first define a correlation quantity that measures a ``similarity" between two models. 
Let
\begin{equation*}
s_p(x,\theta) = \frac{\partial \log p(x|\theta)}{\partial\theta}
\end{equation*}
and 
\begin{equation*}
s_q(x,\phi) = \frac{\partial \log q(x|\phi)}{\partial\phi}
\end{equation*}
be the score functions of $p(x|\theta)$ and $q(x|\phi)$, respectively.
For a given $\theta$, the parameter of the decoder model, $\phi$, is taken to be a function $\phi(\theta)$ of $\theta$ satisfying
\begin{equation}
\label{eq:phi-of-theta}
E[s_q(x,\phi(\theta))|\theta]=0.
\end{equation}
We define the square correlation coefficient $\rho^2_{\theta}$ as
\begin{equation}
\label{eq:rho2}
\rho^2_{\theta} \equiv
\frac{ \mathrm{Cov}[s_p(x,\theta),s_q(x,\phi(\theta))|\theta]^2}
{\mathrm{Var}[s_p(x,\theta)|\theta] \mathrm{Var}[s_q(x,\phi(\theta))|\theta]}
=
\frac{E[s_p(x,\theta)s_q(x,\phi(\theta))|\theta]^2}
{J_{\theta}E[s_q(x,\phi(\theta))^2|\theta]},
\end{equation}
where $\mathrm{Var}[\cdot|\theta]$ and $\mathrm{Cov}[\cdot|\theta]$ represent, respectively, the variance and the covariance with respect to $p(x|\theta)$, and 
$J_{\theta}$ is the Fisher information defined by
\begin{equation*}
J_{\theta} \equiv E[s_p(x,\theta)^2|\theta].
\end{equation*}
Note that we used $E[s_p(x,\theta)|\theta]=0$ in deriving the right-hand side of Eq.(\ref{eq:rho2}).
The square correlation coefficient $\rho_{\theta}^2$ is related to the coefficient of determinant, $R^2$, used in a simple regression analysis \citep{Rawlings98}.

$\rho^2_{\theta}$ has the following geometrical property. 
In a linear space of square integral functions, the inner product and norm are defined to be
\begin{equation*}
\langle s_p, s_q \rangle_{\theta} = E(s_ps_q|\theta),
\end{equation*}
\begin{equation*}
\|s\|_{\theta} = \langle s,s \rangle_{\theta}^{1/2} = E(s^2|\theta)^{1/2}.
\end{equation*}
The square correlation coefficient is then rewritten as
\begin{equation}
\label{eq:geometry}
\rho^2_{\theta} = 
\langle \frac{s_p(x,\theta)}{\| s_p(x,\theta) \|},
\frac{s_q(x,\phi(\theta))}{\| s_q(x,\phi(\theta)) \|} \rangle_{\theta}^2
= cos^2\varphi,
\end{equation}
where $\varphi$ is the angle between $s_p(x,\theta)$ and $s_q(x,\phi(\theta))$ with respect to $\langle , \rangle_{\theta}$.
Thus, $\rho^2_{\theta}=1$ if $s_q(x,\phi(\theta))$ is parallel to $s_p(x,\theta)$, 
while $\rho^2_{\theta}=0$ if $s_q(x,\phi(\theta))$ is orthogonal to $s_p(x,\theta)$.

In the following, we will give two interpretations of $\rho^2_{\theta}$, in terms of statistical inference (Lemma \ref{lem:efficiency}) and information theory (Lemma \ref{lem:information}), which will provide useful insights for translating the meaning of $\rho^2_{\theta}$ into the context of neural decoding. 

\subsubsection{Asymptotic efficiency}\label{sec:efficiency}

Let $x_1,x_2,\ldots,x_n$ be independent and identically distributed random variables from $p(x|\theta)$, and 
$\hat{\phi}_n=\hat{\phi}_n(x_1,x_2,\ldots,x_n)$ be the MLE of $q(x|\phi)$ based on  $x_1,x_2,\ldots,x_n$. 
Then, $\hat{\phi}_n \to \phi(\theta)$ as $n\to\infty$, where $\phi(\theta)$ satisfies Eq.(\ref{eq:phi-of-theta}) \citep{White82}.
For the inference of $\theta$ from $\hat{\phi}_n$, we assume that 
$d\phi(\theta)/d\theta\neq 0$. 
An estimator of $\theta$ would, thus, be transformed from $\hat{\phi}_n$ as  
$\hat{\theta}_n=\phi^{-1}(\hat{\phi}_n)$.
We also assume that $\hat{\theta}_n$ is an unbiased estimator of $\theta$. 
The performance of the unbiased estimator is evaluated by its variance, 
and the ratio of it to its lower bound is called the {\it efficiency} \citep{Schervish95}.  
The following lemma holds under the above conditions. 
\begin{Lemma} \label{lem:efficiency}
$\rho^2_{\theta}$ gives the asymptotic efficiency of $\hat{\theta}_n$.
\end{Lemma}
\begin{Proof}
Under suitable regularity conditions, it is proven that $\hat{\phi}_n$ is asymptotically normal \citep{White82}:
\begin{equation*}
\sqrt{n} (\hat{\phi}_n - \phi(\theta)) \to N(0,v)
\quad \mathrm{in\ distribution},
\end{equation*}
where
\begin{equation}
\label{eq:v}
v = E[s_q(x,\phi(\theta))^2|\theta] 
E\bigg[ \frac{\partial s_q(x,\phi(\theta))}{\partial \phi} \bigg|\theta \bigg]^{-2} ~.
\end{equation}
By the delta method \citep{Schervish95}, we obtain 
\[
\sqrt{n} (\hat{\theta}_n - \theta) \to N(0,v/c^2)
\quad \mathrm{in\ distribution},
\]
where
\begin{equation}
\label{eq:c}
c = \frac{d\phi(\theta)}{d\theta} = 
-E[s_p(x,\theta)s_q(x,\phi(\theta))|\theta]
E\bigg[\frac{\partial s_q(x,\phi(\theta))}{d\phi} \bigg|\theta \bigg]^{-1}
\end{equation}
is derived by differentiating Eq.(\ref{eq:phi-of-theta}) with respect to $\theta$.
Since the lower bound of the asymptotic variance is given by the inverse of the Fisher information (i.e., the Cram\'er-Rao lower bound), 
the asymptotic efficiency is defined by the ratio $c^2J_{\theta}^{-1}/v$.
Using Eqs.(\ref{eq:v}) and (\ref{eq:c}), we obtain $c^2J_{\theta}^{-1}/v = \rho^2_{\theta}$.
\hfill $\Box$
\end{Proof}

\subsubsection{Information-theoretic quantity}\label{sec:information}

We next connect $\rho^2_{\theta}$ to an information-theoretic measure. 
Consider a situation in which a neuron is subjected to 
a stimulus chosen from a probability distribution, $p(\theta)$.
In information theory, the amount of information about the stimulus transferred through a noisy channel is quantified by the mutual information \citep{Cover-Tomas91}:
\begin{eqnarray}
I &=&
-\int p(x) \log p(x) dx 
+ \int\int p(\theta')p(x|\theta')\log p(x|\theta') d\theta' dx.
\label{eq:mutualinfo}
\end{eqnarray}
The amount of information that can be gained by decoding depends on the probability distribution used in a decoder. 
In order to introduce this information, we revisit an information-theoretic interpretation of the mutual information. 
Suppose that the neuron is subjected to a set of stimuli, and consider how many stimuli can be encoded in its response. 
If each stimulus is encoded in a sequence of $n (\gg 1)$ ISIs, the upper bound on the number of stimuli that can be encoded almost error-free is $e^{nI}$. 
In decoding, if the true model, $p(x|\theta)$, is used to build a decoder, 
the upper bound of the number of stimuli that can be decoded almost freely from errors is the same, $e^{nI}$. 
If, on the other hand, the inaccurate model, $q(x|\theta)$, is used, 
then the upper bound is typically smaller, $e^{nI^*}$, 
where $I^* \le I$ was derived in \cite{Merhav-Kaplan-Lapidoth-Shamai94} as 
\begin{equation}
I^* = I^*(\beta^*) =
-\int p(x) \log \int p(\theta') q(x|\phi(\theta'))^{\beta^*}d\theta' dx 
+ \int\int p(\theta')p(x|\theta')\log q(x|\phi(\theta'))^{\beta^*} d\theta' dx,
\label{eq:misinfo}
\end{equation}
with $\beta^*$ being the value that maximizes $I^*(\beta)$.
Thus, the normalized quantity, $I^*/I$, is regarded as an information gain obtained by using $q(x|\phi)$ in decoding.  
See \citet{Latham-Nirenberg05,Oizumi-Ishii-Ishibashi-Hosoya-Okada10} for more details and use of $I^*$ in the context of neural decoding. 
The following lemma connects $I^*/I$ with $\rho^2_{\theta}$.

\begin{Lemma}\label{lem:information}
Suppose that the mean and variance of $p(\theta')$ are given by $\theta$ and $\epsilon^2$, respectively. 
For $\epsilon \ll 1$, the information ratio is given by
\begin{equation}
\label{eq:inforatio}
\frac{I^*}{I} = \rho^2_{\theta} + O(\epsilon).
\end{equation}
\end{Lemma}
\begin{Proof}
For an integrable function, $f(x)$, that is twice differentiable, it follows that 
\begin{equation*}
\int f(x)p(\theta')d\theta' =
f(\theta) + \frac{f''(\theta)}{2}\epsilon^2 + O(\epsilon^3).
\end{equation*}
By using this, we obtain
\begin{eqnarray}
I^*(\beta) =
\beta E[s_p(x,\theta)s_q(x,\phi(\theta))|\theta]\epsilon^2 
- \frac{\beta^2}{2}E[s_q(x,\phi(\theta))^2|\theta] \epsilon^2  + O(\epsilon^3).
\label{eq:misinfo2}
\end{eqnarray}
The optimal $\beta^*$ is obtained by maximizing Eq.(\ref{eq:misinfo2}) with respect to $\beta$ as 
\begin{equation}
\beta^* = 
\frac{E[s_p(x,\theta)s_q(x,\phi(\theta))|\theta]}{E[s_q(x,\phi(\theta))^2|\theta]}
+ O(\epsilon).
\label{eq:optbeta}
\end{equation}
Substituting Eq.(\ref{eq:optbeta}) into Eq.(\ref{eq:misinfo2}) leads to
\begin{equation}
I^* \equiv I^*(\beta^*) =
\frac{E[s_p(x,\theta)s_q(x,\phi(\theta))|\theta]^2 }{2E[s_q(x,\phi(\theta))^2|\theta]}\epsilon^2
+ O(\epsilon^3).
\label{eq:misinfo3}
\end{equation}
In the same manner, the mutual information (\ref{eq:mutualinfo}) is evaluated as
\begin{equation}
I = \frac{1}{2}J_{\theta}\epsilon^2 + O(\epsilon^3).
\label{eq:mutualinfo2}
\end{equation}
From Eqs.(\ref{eq:misinfo3}) and (\ref{eq:mutualinfo2}), we obtain Eq.(\ref{eq:inforatio}).
\hfill $\Box$
\end{Proof}

\subsubsection{Properties of $\rho^2_{\theta}$}\label{sec:properties}

\begin{Lemma} \label{lem:properties}
$\rho^2_{\theta}$ has the following properties:
\begin{enumerate}
\item[i)]
$0 \le \rho^2_{\theta} \le 1$.
\item[ii)]
$\rho^2_{\theta}$ achieves unity when the MLE of $q(x|\phi)$ is a complete sufficient statistic for $\theta$\footnote{
A statistic $\hat{\phi}$ is {\it complete} if for every measurable, real-valued function $f$, 
$E[f(\hat{\phi})|\theta]=0$ for all $\theta \in \Theta$ implies $f(\hat{\phi})=0$ almost surely with respect to $p(x|\theta)$ (denoted by `a.s. [$p_{\theta}$]') for all $\theta$. 
An interpretation of completeness for a sufficient statistic is that it makes the ancillary part of the data independent of $\hat{\phi}$ \citep{Lehmann81}.
}.
\end{enumerate}
\end{Lemma}

\begin{Proof}
i) is obvious from Eq.(\ref{eq:geometry}). 
To prove ii), let $\hat{\phi}=\hat{\phi}(x)$ denote the MLE of $q(x|\phi)$.
Let $f_1(\hat{\phi})$ and $f_2(\hat{\phi})$ be unbiased estimators of $\theta$. 
Then, we have $E[f_1(\hat{\phi})-f_2(\hat{\phi})|\theta]=0$ for all $\theta \in \Theta$.
Since $\hat{\phi}$ is a complete statistic, 
it follows that $f_1(\hat{\phi})=f_2(\hat{\phi})$, a.s. [$p_{\theta}$] for all $\theta$.
Thus, all unbiased estimators of $\theta$, which are functions of $\hat{\phi}$, are equal, a.s. [$p_{\theta}$]. 
Now, suppose that there is an unbiased estimator $f(x)$ of $\theta$ with finite variance, and define 
\begin{equation}
\hat{\theta} = E[f(x)|\theta,\hat{\phi}],
\label{eq:theta_hat}
\end{equation}
which forms an estimator of $\theta$, since $\hat{\phi}$ is sufficient for $\theta$ and thus the conditional expectation given $\hat{\phi}$ does not depend on $\theta$. 
$\hat{\theta}$ is unbiased because 
\begin{equation*}
E(\hat{\theta}|\theta) = E[ E[f(x)|\theta,\hat{\phi}]|\theta]
= E[f(x)|\theta] = \theta.
\end{equation*}
Thus, $\hat{\theta}$ defined by Eq.(\ref{eq:theta_hat}) is equal with the one defined in Lemma \ref{lem:efficiency}, a.s. [$p_{\theta}$]. 
It follows that 
\begin{eqnarray*}
\mathrm{Var}(\hat{\theta}|\theta) &=& 
E[ (\hat{\theta}-\theta)^2 |\theta] \nonumber\\
&=&
E[ (E[f(x)|\theta,\hat{\phi}] - \theta)^2|\theta] \nonumber\\
&=&
E[ E[f(x) - \theta|\theta,\hat{\phi}]^2|\theta] \nonumber\\
&\le&
E[E[(f(x)-\theta)^2|\theta,\hat{\phi}] |\theta] \nonumber\\
&=& 
E[(f(x)-\theta)^2|\theta] \nonumber\\
&=&
\mathrm{Var}[f(x)|\theta], 
\end{eqnarray*}
where the inequality follows from Jensen's inequality. 
Particularly, if we take $f(x)$ to be an asymptotically efficient estimator (e.g., the MLE of $p(x|\theta)$), 
$\mathrm{Var}[f(x)|\theta]$ achieves the lower bound, $J_{\theta}^{-1}$, which completes the proof of ii) because $\rho^2_{\theta}$ gives the asymptotic efficiency of $\hat{\theta}$. 
\hfill $\Box$
\end{Proof}

From the interpretations and properties given in Lemmas \ref{lem:efficiency}, \ref{lem:information} and \ref{lem:properties}, $\rho^2_{\theta}$ can be used as a measure of 
decoding performance of $q(x|\phi)$ when the true model is given by $p(x|\theta)$. 
We say that $q(x|\phi)$ {\it efficiently decodes} $\theta$ if $\rho^2_{\theta}=1$.
If $\rho^2_{\theta}>0$, $\theta$ is {\it asymptotically decodable} with $q(x|\phi)$, whereas if $\rho^2_{\theta}=0$, $\theta$ is {\it not decodable} with $q(x|\phi)$.

\subsection{Results}
\label{sec:neuraldecoding}
By using $\rho^2_{\theta}$ defined in Eq.(\ref{eq:rho2}), 
we now investigate the extent to which the decoders of each scheme decode rate and temporal codes. 

\begin{Theorem}\label{thm:rate-decoding}
In rate encoding, 
if the sample mean is a complete sufficient statistic for $\mu$, the rate decoder efficiently decodes $\theta$ (i.e., $\rho^2_{\theta}=1$ with $q(x|\phi)$ being the exponential distribution). 
\end{Theorem}
\begin{Proof}
Since $\mu(\theta)$ is a one-to-one mapping, the sample mean is sufficient for $\theta$. 
On the other hand, the MLE of the rate parameter of the exponential distribution is given by the sample mean.
Therefore, the theorem follows from Lemma \ref{lem:properties} ii). 
\hfill $\Box$
\end{Proof}

\begin{Theorem}
Let $q(x|\phi)$ be the MI model given by (\ref{eq:mimodel}). 
Either in rate encoding or in temporal encoding, 
\begin{enumerate}
\item[i)]
$\theta$ is efficiently decoded 
(i.e., $\rho^2_{\theta}=1$) if $G(x)$ is a complete sufficient statistic for $\theta$.
\item[ii)]
$\theta$ is asymptotically decodable 
(i.e., $\rho^2_{\theta}>0$) if $\frac{\partial E[G(x)|\theta]}{\partial \theta} \neq 0$.
\end{enumerate}
\end{Theorem}
\begin{Proof}
From Eq.(\ref{eq:mimodel}), the MLE of $q(x|\phi)$ is given by $\hat{\phi}=G(x)^{-1}$. 
Thus, i) follows from Lemma \ref{lem:properties} ii). 
For the proof of ii), we rewrite (\ref{eq:rho2}) as follows.
\begin{eqnarray*}
E[s_p(x,\theta)s_q(x,\phi)|\theta] &=& 
\int \frac{\partial \log p(x|\theta)}{\partial\theta}s_q(x,\phi)p(x|\theta)dx \nonumber\\
&=&
\frac{\partial}{\partial\theta}\int s_q(x,\phi)p(x|\theta)dx \nonumber\\
&=& 
\frac{\partial}{\partial\theta}E[s_q(x,\phi)|\theta] \nonumber\\
&=&
-\frac{\partial}{\partial \theta}E[G(x)|\theta],
\end{eqnarray*}
where we used Eq.(\ref{eq:mimodel}) to obtain the last equation.
Inserting $\phi=\phi(\theta)$ into the above equation leads to 
\begin{equation}
E[s_p(x,\theta)s_q(x,\phi(\theta))|\theta] =
- \frac{\partial}{\partial\theta} E[G(x)|\theta].
\label{eq:enumerator}
\end{equation}
Through direct calculation, we also obtain
\begin{equation}
\label{eq:denominator}
E[s_q(x,\phi(\theta))^2|\theta]  =
E\{(G(x)-E[G(x)|\theta])^2|\theta\}
\equiv \mathrm{Var}[G(x)|\theta]. 
\end{equation}
Substituting Eqs.(\ref{eq:enumerator}) and (\ref{eq:denominator}) into Eq.(\ref{eq:rho2}), $\rho^2_{\theta}$ is written as
\begin{equation}
\label{eq:rho2-mimodel}
\rho^2_{\theta} = 
\frac{ \Big\{ \frac{\partial}{\partial\theta} E[G(x)|\theta] \Big\}^2}
{J_{\theta}  \mathrm{Var}[G(x)|\theta] }~.
\end{equation}
Therefore, $\rho_{\theta}^2 > 0$ holds if $\frac{\partial E[G(x)|\theta]}{\partial \theta} \neq 0$.
\hfill $\Box$
\end{Proof}

The results and their consequences are summarized as follows.
\begin{enumerate}
\item[1)] 
In rate encoding, if the sample mean is a complete sufficient statistic for $\mu$, the rate decoder efficiently decodes the rate code.  
\item[2)] 
If, on the other hand, the sample mean is not sufficient for $\mu$ in rate encoding, but $G(x)$ is chosen so that the value of $\rho^2_{\theta}$ for the temporal decoder is larger than that for the rate decoder, the temporal decoder can decode the rate code with greater efficiency than the rate decoder. 
\item[3)] 
In temporal encoding, if $G(x)$ is chosen so that $\frac{\partial E[G(x)|\theta]}{\partial \theta} \neq 0$, 
the temporal code is asymptotically decodable with the temporal decoder. 
Particularly, if $G(x)$ can be taken to be a complete sufficient statistic for $\theta$, the temporal decoder decodes the temporal code efficiently.
\end{enumerate}

In the following, we will give three examples that illustrate the above consequences. 
We first give an example illustrating consequence 2), where the rate decoder is not efficient for decoding a rate code, and the temporal decoder achieves greater efficiency than the rate decoder. 

\begin{Example}\label{ex:lognormal}
Let $p(x|\mu,\kappa)$ be a log-normal distribution:
\begin{equation}
\label{eq:lognormal}
p(x|\mu,\kappa) = \frac{1}{x\sqrt{2\pi\kappa}}
\exp\bigg[ -\frac{(\log\frac{x}{\mu}+\frac{\kappa}{2})^2}{2\kappa} \bigg] .
\end{equation}
See \cite{Levine91} for modeling the stochastic nature of ISIs with the log-normal distribution. 
Suppose that the stimulus is encoded in $\mu$, i.e., rate encoding. 
The sample mean is not a sufficient statistic for $\mu$ of the distribution, 
which implies that the rate decoder does not decode efficiently. 
Indeed, $\rho^2_{\theta}$ for the rate decoder is derived in Appendix \ref{app:cor_lognormal} as 
\begin{equation}
\label{eq:cor_lognormal}
\rho^2_{\theta} = \frac{\kappa}{e^{\kappa}-1}.
\end{equation}
$\rho^2_{\theta}\to0$ if $\kappa \to \infty$, as the distribution becomes more skewed and has a longer right-hand tail.

Instead of the rate decoder, consider using the temporal decoder with the MI model's recovery function being 
\begin{equation}
\label{eq:refractoryfactor}
g(x) = \frac{(\alpha x/\tau)^{\alpha-1}e^{-\alpha x/\tau}}{\Gamma(\alpha,\alpha x/\tau)},
\end{equation}
where $\Gamma(\alpha, z)$ is the incomplete gamma function:
\begin{equation*}
\Gamma(\alpha,z)= \int_z^{\infty} t^{\alpha-1}e^{-t}dt.
\end{equation*}
In Eq.(\ref{eq:refractoryfactor}), $\alpha (>0)$ determines the shape of $g(x)$ (i.e., $g(x) \sim x^{\alpha-1}$ near $x=0$), and 
$\tau$ represents the correlation timescale between successive spikes. 
Figure~\ref{fig:result}(a) depicts the shape of $g(x)$ for several values of $\alpha$.
It is shown in Appendix \ref{app:temp-logn} that for each $\kappa>0$, the temporal decoder with the recovery function (\ref{eq:refractoryfactor}) achieves $\rho^2_{\theta} \approx 1$ as closely as possible by taking $\tau$ to be large enough and $\alpha$ to be small enough, 
because the sufficient statistic $G(x)$ for the parameter $\phi$ of the MI model approximates to $\log x$, which is a sufficient statistic for the mean parameter of the log-normal distribution. 
\hfill $\Box$
\end{Example}

\begin{figure}[t]
\hfill
\begin{center}
\includegraphics[width=0.95\textwidth]{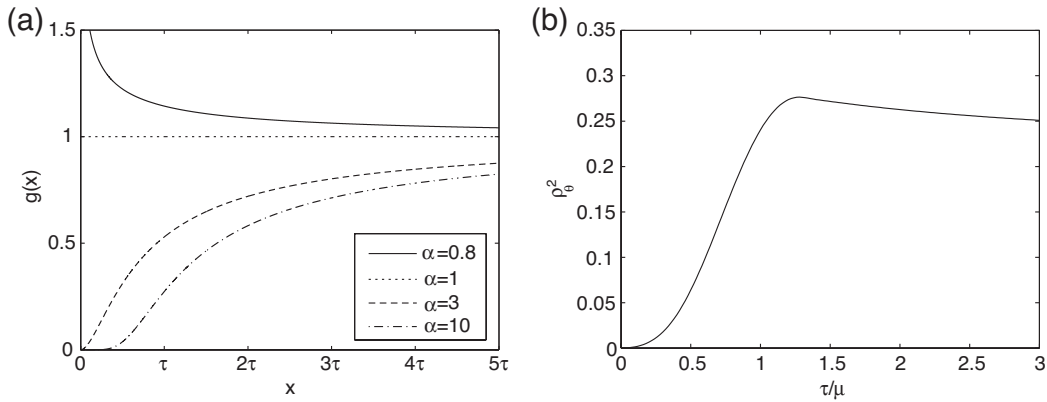}
\end{center}
\caption{
(a) The shape of the recovery function (\ref{eq:refractoryfactor}) for $\alpha=0.8$, 1, 3 and 10. 
(b) $\rho_{\theta}^2$ of temporal decoding as a function of $\tau/\mu$ in Example \ref{ex:correlationcode}.
The value of the shape parameter of the gamma distribution was taken to be $\kappa=5$. 
$\rho^2_{\theta}$ reaches its maximum when $\mu \approx \tau$.
}
\label{fig:result} 
\end{figure}

The next example illustrates consequence 1), i.e., a situation in which the sample mean is sufficient for the mean parameter. 

\begin{Example}\label{eq:gamma-exp}
Suppose that  $p(x|\mu,\kappa)$ is a gamma distribution with the mean $\mu$ and the shape parameter $\kappa$:
\begin{equation}
\label{eq:gammadist}
p(x|\mu,\kappa) = \frac{\kappa^{\kappa}x^{\kappa-1}e^{-\kappa x/\mu}}{\mu^{\kappa}\Gamma(\kappa)},
\end{equation}
where $\Gamma(\kappa)$ is the gamma function. 
The gamma distribution has been used to describe the stochastic nature of ISIs, and its information-theoretic properties have been studied (\cite{Ikeda09} and references therein).
Also, suppose that the stimulus is mapped onto $\mu$ (i.e., rate encoding).
It is easy to  see that the sample mean is a complete sufficient statistic for $\mu$, and thus 
the rate decoder efficiently decodes the stimulus ($\rho^2_{\theta}=1$), regardless of the value of $\kappa$. 
Note that the variance of the sample mean achieves the Cram\'er-Rao lower bound even with a finite sample size, because the gamma distribution is an exponential family distribution \citep{Schervish95}. Thus, neither the temporal decoder nor the gamma distribution (i.e., the true model) is necessary for efficient decoding even with a finite sample size. 
\hfill $\Box$
\end{Example}

The last example illustrates consequence 3).

\begin{Example}\label{ex:correlationcode}
Consider that the true ISI distribution is given to be the gamma distribution (\ref{eq:gammadist}), and that the stimulus is encoded in $\kappa$ (i.e., temporal encoding). 
For temporal decoding, let us take the recovery function of the MI model to be Eq.(\ref{eq:refractoryfactor}). 
From a direct calculation (Appendix \ref{app:temp-gamma}), $\rho^2_{\theta}$ is expressed as 
\begin{equation}
\label{eq:temp-gamma}
\rho^2_{\theta} = 
\frac{ \big\{ \frac{\partial}{\partial\kappa}E[\log\Gamma(\alpha,\alpha\frac{\mu}{\tau}x)|\kappa] \big\}^2 }
{J_{\kappa}\mathrm{Var}[\log\Gamma(\alpha,\alpha\frac{\mu}{\tau}x)|\kappa]}, 
\end{equation}
where $E[\cdot|\kappa]$ and $\mathrm{Var}[\cdot|\kappa]$ are taken with respect to $p(x|\mu=1,\kappa)$. 
Note that $\rho^2_{\theta}$ is a function of the dimensionless parameter, $\mu/\tau$.
$\rho^2_{\theta}$ was numerically computed for each value of the parameters, $(\kappa, \mu/\tau)$. 
The value of $\alpha$ was taken so as to maximize $\rho^2_{\theta}$ for each value of parameters.
Figure~\ref{fig:result}(b) depicts $\rho^2_{\theta}$ as a function of $\tau/\mu$. 
It is seen from this figure that $\rho^2_{\theta}$ takes its maximum near $\tau/\mu\approx 1$, which indicates that the MI model decodes best when the mean ISI of the true model, $\mu$, matches the correlation timescale of the MI model, $\tau$. 
\hfill $\Box$
\end{Example}

\section{Discussion}\label{sec:discussion}

Our main results are summarized as follows. 
First, the rate decoder efficiently decodes rate codes if and only if the sample mean is a sufficient statistic for the mean parameter of the true model. 
Second, the temporal decoder improves on the performance of the rate decoder by 
a) decoding temporal codes that the rate decoder fails to read, and 
b) achieving greater efficiency in decoding certain rate codes.

These results suggest that rate codes in stationary spike trains, which are defined as the mapping from the stimulus to the mean firing rate, can further be divided into two subcategories 
when the concept of sufficiency is taken into consideration:  
one is a ``strong" rate code, in which the sample mean is a sufficient statistic for decoding, and the other is a ``weak" rate code, in which the sample mean is not sufficient. 
We should notice that spike count decoding matches the strong form of rate encoding, but not weak form. 

How can decoding results inform us whether or not rate coding is being used?
In order to answer this question in the context of neuronal data analysis, 
one may decode the stimulus with rate and temporal decoders, and compare their decoding performances \citep{Jacobs-et-al09}. 
This procedure tells us whether or not the sample mean is sufficient for decoding the stimulus: if the rate decoder performs as well as the temporal decoder, then the sample mean is sufficient; if it does not, then the sample mean is not sufficient.
In terms of the original question of whether rate coding is being used, only in the former case can we translate the decoding result into ``strong" rate encoding; 
in the latter case, we cannot conclude which scheme, ``weak" rate encoding or temporal encoding, is being used.

The key quantity in our theoretical analysis is the square correlation coefficient, $\rho^2_{\theta}$, which quantifies neural decoding performance. 
It is worth pointing out that the unnormalized quantity of $\rho^2_{\theta}$:
\begin{equation*}
J^*_{\theta} \equiv \rho^2_{\theta} J_{\theta} = 
\frac{E[s_p(x,\theta)s_q(x,\phi(\theta))|\theta]^2}
{E[s_q(x,\phi(\theta))^2|\theta]},
\end{equation*}
can be regarded as a generalization of the Fisher information, $J_{\theta}$, in the sense that 
$J^*_{\theta}$ becomes $J_{\theta}$ if $q(x|\phi)=p(x|\theta)$. 
$J^*_{\theta}$ has similar properties to $J_{\theta}$; 
(i) ${J^*_{\theta}}^{-1}$ gives the asymptotic variance of the MLE of $q(x|\phi)$ (Lemma \ref{lem:efficiency}) as ${J_{\theta}}^{-1}$ gives that of $p(x|\theta)$ \citep{Schervish95}, 
and 
(ii) $J^*_{\theta}$ appears in the leading term of the information, $I^*$, of the decoder with $q(x|\phi)$ (Lemma \ref{lem:information}), as $J_{\theta}$ does in the mutual information with the limit of small input power \citep{Kostal10}. 
As $J_{\theta}$ has been used to measure encoding accuracy 
\citep[for review, see][chap. 3]{Dayan-Abbott01},	
$J^*_{\theta}$ is used to measure the performance of neural decoders. 

It must be noted that our analysis is based on asymptotic theory, which assumes a large sample size. 
The inverse of the Fisher information and its generalization, $J_{\theta}^*$, give the lower bounds of the variance of unbiased estimators, but generally do not correspond to the mean squared error of the estimators with a finite sample size, except for special cases of exponential family distributions. 
Thus, the results based on asymptotic analysis may not be justified for non-asymptotic cases. 
(\cite{Bethge02} examined this point in the context of population coding.)
Especially, decoding using the ``wrong" model may severely compromise the accuracy of decoding in non-asymptotic cases. 
One therefore has to check carefully whether analysis using $\rho^2_{\theta}$ provides correct results in terms of minimum mean squared error 
when the asymptotic results are translated into non-asymptotic cases. 

Our simple setting of stationary and renewal assumptions does not account for two aspects of neuronal spikes that are relevant for neural coding. 
First, actual spike trains exhibit nonstationarity due to both, the dynamics of the stimulus and the nature of the neural encoding processes such as adaptation. 
Rate encoding for this case is generalized to the scheme in which the stimulus is mapped onto a time-dependent firing rate, or, the marginal intensity function. 
Then the question we would like to address is  
whether reasonable estimates of the firing rate (e.g., based on spline models or histograms), are asymptotically sufficient for decoding the stimulus, 
which may require more mathematically careful treatment to be proven. 
Second, higher-order serial dependencies in sequences of ISIs, for which the MI model (\ref{eq:mi}) can not account, would certainly be relevant for neural coding. 
Accordingly, temporal encoding is generalized to the scheme in which the stimulus is  mapped onto the higher-order serial dependencies. 
For temporal decoding, the MI model can be generalized by taking the recovery function to depend on the whole spiking history, rather than simply on the last spike.
Taking into consideration these two extensions, we suspect that our results summarized at the beginning of the Discussion still hold. 
It would be interesting to examine the relation between encoding and decoding 
in a more realistic setting, for instance, with biophysically realistic neuron models.

\appendix
\section{Appendix: details of derivations}
\subsection{Derivation of equation (\ref{eq:cor_lognormal})}\label{app:cor_lognormal}
Taking the parameter $\mu=\mu(\theta)$ and inserting $G(x)=x$ into (\ref{eq:rho2-mimodel}), $\rho^2_{\theta}$ for the rate becomes
\begin{equation*}
\rho^2_{\theta} = 
\frac{ \Big\{ \frac{\partial}{\partial\mu} E(x|\theta) \Big\}^2}
{J_{\mu}  \mathrm{Var}(x|\theta) }~.
\end{equation*}
For the log-normal distribution (\ref{eq:lognormal}), we have 
$E(x|\theta)=\mu$, 
$\mathrm{Var}(x|\theta)= \mu^2(e^{\kappa}-1)$, 
and
\begin{equation*}
J_{\mu} = - E\bigg[ \frac{\partial^2}{\partial\mu^2}\log p(x|\mu,\kappa) \bigg|\theta\bigg] 
= \frac{1}{\kappa\mu^2}.
\end{equation*}
Using these, we obtain Eq.(\ref{eq:cor_lognormal}).

\subsection{Temporal decoding for the log-normal distribution}\label{app:temp-logn}
Here, we show that the temporal decoder with recovery function (\ref{eq:refractoryfactor}) can achieve $\rho^2_{\theta}\approx 1$ as closely as possible in Example \ref{ex:lognormal}. 
Taking the parameter $\mu=\mu(\theta)$ in (\ref{eq:rho2-mimodel}), we have
\begin{equation*}
\rho^2_{\theta} = 
\frac{ \Big\{ \frac{\partial}{\partial\mu} E[G(x)|\theta] \Big\}^2}
{J_{\mu}  \mathrm{Var}[G(x)|\theta] }~,
\end{equation*}
where 
\begin{eqnarray*}
G(x) &=& \frac{\tau}{\alpha}\bigg\{
\log\Gamma(\alpha) - \log\Gamma\Big( \alpha,\frac{\alpha x}{\tau} \Big)
\bigg\} \nonumber\\
 &=&
 \log\Gamma(\alpha) - \frac{\alpha^{\alpha-1}x^{\alpha}}{\Gamma(\alpha)\tau^{\alpha}} + O(\tau^{-\alpha-1}), 
\end{eqnarray*}
for $\tau \gg 1$.
Then, 
\begin{equation*}
\frac{\partial E[G(x)|\theta]}{\partial\mu} 
= -\frac{\alpha^{\alpha-1}}{\Gamma(\alpha)\tau^{\alpha}}
\frac{\partial E(x^{\alpha}|\theta) }{\partial\mu} + O(\tau^{-\alpha-1}).
\end{equation*}
A similar calculation leads to
\begin{equation*}
\mathrm{Var}[G(x)|\theta] =
\bigg(  \frac{\alpha^{\alpha-1}}{\Gamma(\alpha)\tau^{\alpha}} \bigg)^2
\mathrm{Var}(x^{\alpha}|\theta) + O(\tau^{-2\alpha-1}).
\end{equation*}
For the log-normal distribution (\ref{eq:lognormal}), we also have 
$J_{\mu}=1/(\kappa\mu^2)$ and 
$E(x^m|\theta)=\mu^m e^{\kappa(m-1)m/2}$, $m>0$. 
Thus, we obtain
\begin{equation*}
\rho^2_{\theta} = 
\frac{\kappa\alpha^2}{e^{\kappa\alpha^2}-1} + O(\tau^{-1}).
\end{equation*}
Therefore, 
$\lim_{\tau\to\infty,\alpha\to0}\rho^2_{\theta} =1$, that is, 
we can achieve $\rho^2_{\theta}\approx 1$ as closely as possible 
by taking $\tau$ to be large enough and $\alpha$ to be small enough.

\subsection{Derivation of equation (\ref{eq:temp-gamma})}\label{app:temp-gamma}

Eq.(\ref{eq:refractoryfactor}) is rewritten as 
\begin{equation*}
g(x) = -\frac{ \frac{\tau}{\alpha}\frac{\partial\Gamma(\alpha,\alpha x/\tau)}{\partial x} }{\Gamma(\alpha,\alpha x/\tau)}
= -\frac{\tau}{\alpha} \frac{\partial}{\partial x} \log \Gamma(\alpha,\alpha x/\tau).
\end{equation*}
Then, we get
\begin{equation*}
G(x) = \int_0^xg(u)du = \frac{\tau}{\alpha}[ \log\Gamma(\alpha) - \log \Gamma(\alpha,\alpha x/\tau)],
\end{equation*}
where we used $\Gamma(\alpha,0)=\Gamma(\alpha)$.
Taking $\kappa=\kappa(\theta)$ in Eq.(\ref{eq:rho2-mimodel}), we obtain
\begin{equation*}
\rho^2_{\theta} = \frac{ \Big\{ \frac{\partial}{\partial\kappa} E[G(x)|\theta] \Big\}^2}
{J_{\kappa}  \mathrm{Var}[G(x)|\theta] } 
=
\frac{ \big\{ \frac{\partial}{\partial\kappa}E[\log\Gamma(\alpha,\alpha x/\tau)|\theta] \big\}^2 }
{J_{\kappa}\mathrm{Var}[\log\Gamma(\alpha,\alpha x/\tau)|\theta]}.
\end{equation*}
Thus, the scaling property of the gamma distribution leads to Eq.(\ref{eq:temp-gamma}).


\end{document}